# Simulation of crack induced nonlinear elasticity using the combined finite-discrete element method

Ke Gao[a*], Esteban Rougier[a], Robert A. Guyer[a,b], Zhou Lei[a], and Paul A. Johnson[a]

[a] *Geophysics, Los Alamos National Laboratory, New Mexico, USA*
[b] *Department of Physics, University of Nevada at Reno, Nevada, USA*

## Abstract

Numerical simulation of nonlinear elastic wave propagation in solids with cracks is indispensable for decoding the complicated mechanisms associated with the nonlinear ultrasonic techniques in Non-Destructive Testing (NDT). Here, we introduce a two-dimensional implementation of the combined finite-discrete element method (FDEM), which merges the finite element method (FEM) and the discrete element method (DEM), to explicitly simulate the crack induced nonlinear elasticity in solids with both horizontal and inclined cracks. In the FDEM model, the solid is discretized into finite elements to capture the wave propagation in the bulk material, and the finite elements along the two sides of the crack also behave as discrete elements to track the normal and tangential interactions between crack surfaces. The simulation results show that for cracked models, nonlinear elasticity is generated only when the excitation amplitude is large enough to trigger the contact between crack surfaces, and the nonlinear behavior is very sensitive to the crack surface contact. The simulations reveal the influence of normal and tangential contact on the nonlinear elasticity generation. Moreover, the results demonstrate the capabilities of FDEM for decoding the causality of nonlinear elasticity in cracked solid and its potential to assist in Non-Destructive Testing (NDT).

## Keywords

Nonlinear elasticity; Crack; Wave propagation; Non-Destructive Testing (NDT); Combined finite-discrete element method (FDEM);

---

[*] Corresponding author: K. Gao (kegao@lanl.gov)

# 1. Introduction

Non-Destructive Testing (NDT) for detection and quantification of defects in solids (e.g. crack, delamination, debonding, pore and inter-granular contact) is of significant industrial and academic importance in many areas [1-5]. Probing and imaging applying ultrasonic waves is a leading tool for such applications and there is a need for continued development of robust ultrasonic techniques in NDT [2,3]. Among the many ultrasonic wave applications in NDT, a major class of them employ the principles of wave reflection, transmission or scattering. While extremely useful, these linear ultrasonic techniques are less capable of accurately detecting contact-type defects, and are also less sensitive to micro or closed cracks [2-4]. Since defects can behave nonlinearly under sufficient excitation, the Nonlinear Elastic Wave Spectroscopy (NEWS) methods have shown remarkable potential for defect detection and characterization [2-4,6-10]. The nonlinear methods often involve exciting a solid with an ultrasonic signal of certain frequency and generating an output frequency spectrum consisting of harmonics and subharmonics of the exciting frequency. These effects, often referred to as Contact Acoustic Nonlinearity (CAN) [10], are mainly induced by clapping and frictional contacts of the defects [2,3,11]. Because of their high sensitivity, nonlinear ultrasonic approaches have seen an increasing interest in the NDT community over the past decades [1,2,4,9,12,13].

A large number of experimental studies have been conducted using NEWS techniques for various types of defects detection in solid materials such as composite plates, metals, concretes and rocks, and have demonstrated successes in the field of NDT [6-8,14-21]. However, to date, the underlying microscopic mechanism of nonlinear techniques for defect detection is still poorly understood [4]. Numerical simulations, which are capable of providing detailed analyses of the nonlinear behavior at a level of spatial and temporal resolution not accessible experimentally, are therefore necessary for decoding the complicated mechanism associated with the nonlinear ultrasonic techniques. In addition to the ease of implementation, the numerical approaches are indispensable also because they can provide dedicated comparisons of nonlinear indicators with experimental results, and thus link measured macroscopic events to defect internal parameters (both physical and geometric) and in this way, a complete characterization of the defects can be achieved [2].

Simulation of ultrasonic wave propagation in solids with defects by considering the nonlinearity introduced by them is challenging, and has been the object of study during the last decades. The numerical approaches used in this regard mainly include Finite-Difference Time-Domain (FDTD) and Finite Element Method (FEM). For example, Sarens *et al.* [22] implemented a three-dimensional finite difference, staggered grid simulation to model the contact nonlinear acoustic generation in a composite plate containing an artificial defect; Marhenke *et al.* [23] used FDTD simulations to cross-validate the simulated interference effects resulting from multiple ultrasonic reflections within the delamination layers with the laboratory

*2/27*

experiments. In general, the FDTD is easy to implement; however, it has many restrictions, e.g. the defects in the FDTD simulations are usually restricted to rectangular shapes [2]. As a more flexible alternative, FEM is widely used in crack-wave interaction simulations. In particular, Kawashima *et al.* [24] used a FEM model to study CAN in which Rayleigh waves were employed to detect surface cracks; Blanloeuil *et al.* [12,25] studied the nonlinear scattering of ultrasonic waves by closed cracks subjected to CAN; to investigate the clapping and friction induced nonlinearity in solids containing cracks, Van Den Abeele and colleagues [2,3,11,26] implemented a series of comprehensive normal and tangential constitutive models into FEM to control the nonlinear behavior of crack surfaces. Among the many numerical simulations, some of them use hypothetical defects in which artificial nonlinear stress-strain relations are introduced into special elements to represent defects [24]; others employ physical defects by splitting the computational nodes along defects, and then the normal and tangential contact stresses, which are calculated based on the relative distances between the corresponding Gauss points located on the defects surfaces, are applied to the same Gauss point pairs as boundary conditions for the bulk material simulation [2-5,11,12,22,23,25-29]. The former only captures the defect behavior in an approximate manner; the latter may be difficult to explicitly realize complicated scenarios such as defects with irregular shapes and especially, the interactions between many defects of different types.

From a computational mechanics point of view, a solid with defects is essentially a combination of continua (bulk material) and discontinua (interaction between defect surfaces). Considering this, a numerical tool that has the capability of handling continua and discontinua simultaneously would be helpful. Fortunately, a recently developed numerical method – the combined finite-discrete element method (FDEM) [30-33], which merges finite element-based analysis of continua with discrete element-based transient dynamics, contact detection and contact interaction solutions of discontinua, provides a natural solution for such simulation. To date, a systematic application of FDEM in NDT is not available in the literature. The goal of this paper is to introduce FDEM to the NDT community, and to demonstrate its power on simple problems.

A simple FDEM realization of a solid plate with crack is presented in Fig. 1 where the solid (excluding the crack since it is considered as void) is discretized into finite elements to capture the motion and deformation of the bulk material, and the finite elements along the two sides of the crack also behave as discrete elements to track the normal and tangential interactions between crack surfaces. By employing FDEM, the system can be explicitly described and particularly, the contacts along the sides of the defects can be uniformly processed using well-developed discrete element method (DEM)-based algorithms.

The focus of the current work is to demonstrate the applicability of FDEM to the simulation of crack induced nonlinear elasticity in solids, and to present it as another alternative for NDT numerical based analysis. A comparison of the simulated results with laboratory experiments is beyond the scope of the present paper and thus it is left for future work. In the following sections, we first provide a brief introduction to the theories of FDEM. Then we illustrate the numerical model setup and present how the normal (clapping) and tangential (friction) contact may influence the nonlinear behavior of a cracked solid. The applicability of FDEM for NDT simulation is demonstrated, and the corresponding conclusions are drawn.

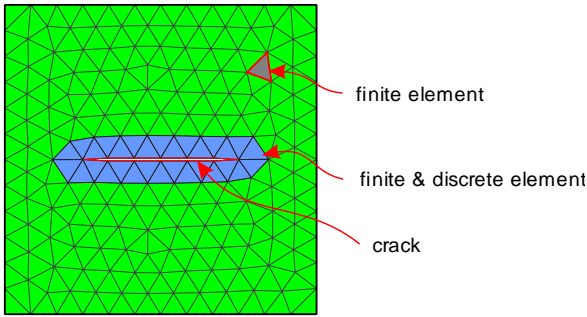

Fig. 1. Numerical representation of a solid plate with a crack using FDEM. The solid is discretized into finite elements to capture the bulk material behavior such as wave propagation. Note that the two crack surfaces do not share nodes, and thus both closed and open cracks can be explicitly represented by altering the crack aperture. The finite elements along the two sides of the crack also behave as discrete elements to simulate the normal and tangential interactions between crack surfaces.

## 2. The combined finite-discrete element method (FDEM)

FDEM was originally developed by Munjiza in the early 1990s to simulate the material transition from continuum to discontinuum [33]. The essence of this method is to merge the algorithmic advantages of DEM with those of the FEM. The theory of the FDEM can be broken down into the following parts: governing equations, finite strain-based formulation for deformation description, contact detection, and contact interaction algorithms [34-36].

*2.1. Governing equations*

The general governing equation of the FDEM is [30]

$$\mathbf{M}\ddot{\mathbf{x}} + \mathbf{C}\dot{\mathbf{x}} = \mathbf{f}, \qquad (1)$$

where $\mathbf{M}$ is the lumped mass matrix, $\mathbf{C}$ is the damping matrix, $\mathbf{x}$ is the displacement vector, and $\mathbf{f}$ is the equivalent force vector acting on each node which includes all forces existing in the system such as the forces due to material deformation and contact forces between solid elements. An explicit time integration scheme based on a central difference method is employed to solve Eq. (1) with respect to time to obtain the transient evolution of the system.

## 2.2. Finite strain based formulation

In FDEM, deformation of the finite elements according to the applied load is described by a multiplicative decomposition-based formulation [32]. This framework allows for a uniform solution for both isotropic and general anisotropic materials [34]. Moreover, volumetric locking due to the lower order finite element implementations can be eliminated by using a selective stress integration scheme.

## 2.3. Contact detection

The finite elements located on the model boundaries (both external and internal) also act as deformable discrete elements. The contact detection between discrete elements is conducted using the MRCK (Munjiza-Rougier-Carney-Knight) algorithm which is based on the decomposition of the simulation space into identical square (two-dimensional) or cubical (three-dimensional) search cells [31,37]. Consider that for any two given elements, one called the contactor and the other one the target, both are mapped onto search cells according to their current position. The goal of the contact search process is to determine whether the contactor and the target share at least one cell. After processing the contact detection, a list that contains all the pairs of elements potentially in contact is established and sent for contact interaction processing. It is worth noting that the MRCK contact detection algorithm is very efficient which is demonstrated by a CPU time proportional to the total number of contact couples, and it is applicable to systems consisting of many bodies of different shapes and sizes [37].

## 2.4. Contact interaction

Contact interaction is critical to wave-crack interaction since it controls the overall nonlinear behavior of the cracked solid. When contact couples are identified, a penalty function based contact interaction algorithm is used to calculate the contact forces between contacting elements [30,31]. In the penalty function method, a small penetration or overlap is allowed between elements in contact, which then determines the normal contact force (magnitude and direction) acting on the contacting elements. In the present work, a "triangle to point" [31] contact interaction algorithm is used in which the target triangular element is discretized into a series of points distributed on its edges and each target point is considered as a Gauss integration point through which the distributed contact forces are integrated (Fig. 2).

In terms of the normal contact force calculation, actual contact will not occur unless the target point is located inside of the contactor triangle. The normal contact force is calculated using the following equation obtained through a derivation in which the energy balance is preserved [30,31]:

$$f_N \mathbf{n}_N = A E_p \frac{h}{H} \mathbf{n}_N, \qquad (2)$$

where $A = l_t / n_t$, $l_t$ is the length of the target element edge on which the target point is located, $n_t$ is the number of target points per element edge, $E_p$ is the penalty parameter, $h$ is the distance between target point and the contactor element edge, $H$ is the height of the contactor element associated with the contacting edge, and $\mathbf{n}_N$ is unit vector of the normal contact force (Fig. 2).

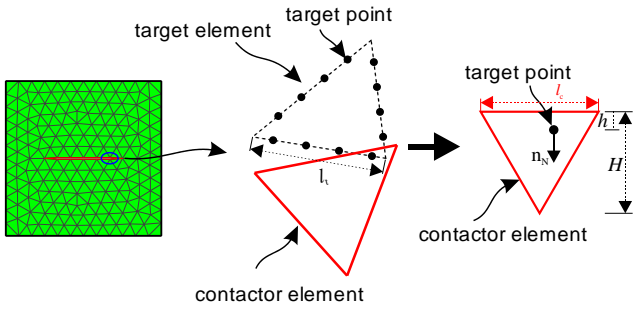

Fig. 2. Schematic of contact interaction and contact force calculation.

The tangential contact force is assessed by means of the friction coefficient and the relative tangential displacement between a target point and contactor element occurring during a time step therefore taking into account the history of contact [38]. The total relative tangential displacement between a target point and a contactor element accumulated at time step $t$, $d_t$, is recorded and used to calculate $d_{t+1}$:

$$d_{t+1} \mathbf{n}_T = \left( d_t + v_r^t \Delta t \right) \mathbf{n}_T, \qquad (3)$$

where $\mathbf{n}_T$ is the unit vector of tangential force, $v_r^t$ is the relative tangential velocity between the target point and contactor element at time step $t$ and $\Delta t$ is the time step. The total tangential force at time step $t$ is calculated by

$$f_T^t \mathbf{n}_T = \left[ \min\left( \mu f_N, A E_p \frac{d_t}{l_c} \right) \right] \mathbf{n}_T, \qquad (4)$$

where $l_c$ is the length of the contactor element edge associated with tangential contact and $\mu$ is the frictional coefficient. The calculated normal and tangential contact forces are distributed among the nodes of contactor element and, in a similar manner, the opposite forces are applied to the target point and distributed among the nodes of the target element.

It is beyond the scope of present paper to provide detailed descriptions of the above principle. However, details of these can be found in several FDEM books [30-32]. FDEM allows explicit geometric and mechanical realization of solids with various types of defects including cracks (open or closed, with flat, bending or rough surfaces), pores and debondings of different shapes, as well as the interactions between them. For the simulations of such problem involving both continua and discontinua, FDEM is superior to both pure FEM and DEM. Since its inception [33], FDEM has proven its computational efficiency and reliability and has been extensively used in a wide range of endeavors in both industry and academia, such as stress heterogeneity [39,40], permeability [41,42], acoustic emission [43] and hydraulic fracturing [44] in rock masses, tunneling [45,46], block caving [47,48] and rock blasting [49] in rock mechanics, red blood cell aggregation in medicine [50], masonry wall stability [51], coastal protection [52], granular fault evolution [53] and shell structure fracturing [54]. Additionally, benefitting from the recent implementation of a large-strain large-rotation formulation and grand scale parallelization in FDEM by the Los Alamos National Laboratory [32,55], the FDEM software package – HOSS (Hybrid Optimization Software Suite) [56,57] – offers a powerful tool to study the crack induced nonlinear elasticity in solid material.

## 3. FDEM simulation examples

As an introductory illustrative application of the use of FDEM for crack induced nonlinearity simulations, two-dimensional rectangle models with a single crack are excited using compressive sinusoidal waves. We first demonstrate the model setup, then a set of simulations using different combinations of crack aperture and excitation amplitudes are presented for the model with a single horizontal crack to examine the applicability of FDEM for this type of applications. Finally, a more detailed exploration of the clapping and friction induced nonlinear elasticity is conducted for models with both horizontal and inclined cracks.

*3.1. Model setup*

Fig. 3 presents the geometry and mesh of the models used in the simulations. The model has a width $W = 25$ mm and height $H = 50$ mm, and two-dimensional plane stress conditions are assumed. A crack with a length $l = 2/3W$ and aperture $a$ is placed in the center of the plate. When building the geometry, the inside of the crack is treated as void, thus the crack aperture is explicitly realized. For the convenience of comparing the normal and shear forces, the friction coefficient used between the internal crack surfaces for contact interaction calculation is set to one. The plate has a density of 2600 kg/m$^3$, Young's modulus of 10 GPa, and Poisson's ratio of 0.25. In terms of the penalty parameter between contacts, theoretically it should be infinity in order to avoid penetration between contacting elements; however, a large penalty parameter will yield a significantly small time step. Recent studies show that, in general, a penalty parameter that is about 1-2 orders of magnitude larger than the Young's modulus can ensure the computational correctness [58]. By compromising between achieving the correct elastic response between contacting elements and maximizing the time step size to reduce the overall computational expense, a penalty parameter 50 times of the plate's Young's modulus, i.e. 500 GPa, is used.

The model consists of 12,894 three-node constant-strain triangular elements and they are approximately uniformly meshed with an average element size of around 0.5 mm. This mesh size is carefully chosen after several trials and ensuring that the total element number is sufficient to precisely capture the wave propagation while also assuring the model is not too computationally expensive. In the current simulation only flat and parallel crack surfaces are used since through these, the crack aperture can be precisely controlled and it is easier to monitor which part of the crack has the most contact when subjected to certain excitation waves. Therefore, the crack used here has a uniform aperture throughout, except at the crack tips, where chamfered corners are used, which occupy only two element size (~1 mm) (see inset of Fig. 3a). A purely linear elastic stress-strain relationship is employed for the elements to avoid introducing any nonlinear source from the bulk material.

A compressive sinusoidal wave excitation with a frequency $f = 100$ kHz is acting uniformly at the bottom boundary of the plate as a velocity boundary condition in the *y*-direction in the form of

$$v_y(t) = A_{v_y} \cdot \cos(2\pi ft), \tag{5}$$

where $A_{v_y}$ is the velocity excitation amplitude. The remaining boundaries in the model are left as free. The velocity of this input wave (Eq. (5)) is obtained by firstly introducing a prescribed strain in *y*-direction ($\varepsilon_y$), which works as a maximum strain intended to excite the model. Then the amplitude of the displacement in the *y*-direction is given by

$$A_{d_y} = H \cdot \varepsilon_y, \tag{6}$$

which corresponds to a sinusoidal wave in terms of *y*-displacement of

$$d_y(t) = A_{d_y} \cdot \sin(2\pi ft). \tag{7}$$

Taking the derivative of $d_y(t)$ with respect to time *t*, the velocity input wave equation in Eq. (5) is obtained, thus we have the velocity amplitude of the input excitation of

$$A_{v_y} = 2\pi f \cdot A_{d_y}, \tag{8}$$

and the velocity input excitation wave of

$$v_y(t) = 2\pi f \cdot H \cdot \varepsilon_y \cdot \cos(2\pi ft). \tag{9}$$

The above gives the relationship between the prescribed strain $\varepsilon_y$, displacement excitation wave $d_y(t)$ and velocity excitation wave $v_y(t)$, respectively. In the following sections, one of the three will be used to describe the input compressive excitation wave, depending on the type of output data that to be analyzed. The other two can be obtained accordingly using the above relationships and may also be provided for reference.

In addition to the horizontal crack where the normal contact is dominant, a crack inclined by 30º (Fig. 3b) is also used to further explore the effect of both normal and tangential contacts on the generation of nonlinear effects. The parameters used in the simulations give a wave length of 19.6 mm, which is approximately equal to 40 element length. The time step is 4.0e-9 s, which corresponds to 2500 time steps per wave cycle. According to the previous investigations conducted using FEM [2], the parameters and meshes used here can guarantee the convergence of the model and will thus produce stable and accurate solutions. The models are run on a parallel cluster utilizing 36 processors and each model has been excited for 52 ms simulation time. A "sensor" point is positioned 5 mm above the model center to track the motion of the plate. The resulting *y*-velocity at the sensor point is exported every 50 time steps (i.e. 2.0e-7 s) and the data between 48 ms and 52 ms are extracted for Fast Fourier transform (FFT) analysis to interpret the nonlinear response of the solid with respect to wave excitation.

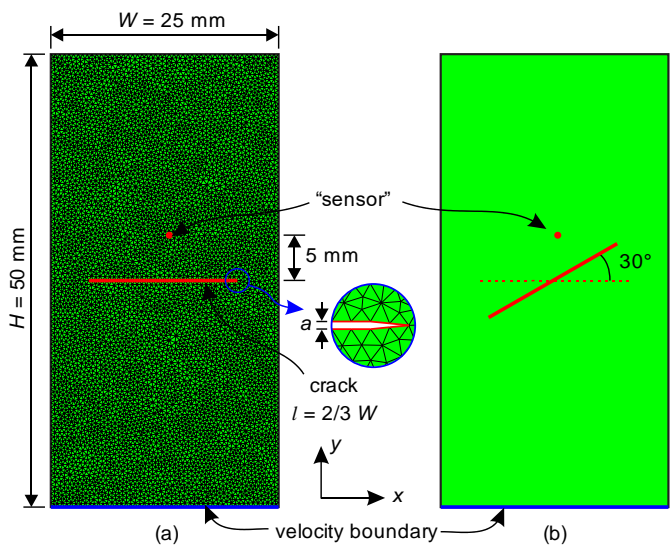

Fig. 3. Illustration of the FDEM model of a solid plate with a single crack: (a) horizontal crack located in the center of the plate and the corresponding mesh; (b) crack inclined 30º with respect to the horizontal direction and located in the plate center.

*3.2. The capability of FDEM for nonlinear elasticity phenomenon acquisition*

To examine the capability of FDEM for capturing nonlinear phenomena induced by the presence of cracks and to explore the relationship between crack aperture and harmonics generation, using the model with horizontal crack (Fig. 3a), we have designed a series of scenarios in which different combinations of crack apertures and excitation amplitudes are tested. As can be seen from Fig. 4, three models with increasing crack apertures of $a = 0$ (closed crack), 5.0e-5 mm and 5.0e-3 mm, respectively, are created. Then three increasing prescribed strains $\varepsilon_y$ = 1.0e-8, 1.0e-6 and 1.0e-4, corresponding to excitation waves with displacement amplitudes of $A_{d_y}$ = 5.0e-7 mm, 5.0e-5 mm and 5.0e-3 mm (based on Eq. (6)), respectively, are used to load the models. Here, only the horizontal crack is used since, compared with the inclined crack, it is relatively easier to foretell the range of crack aperture change when the models are excited using compressive waves at the bottom boundary, and thus it is possible to predict in which scenario the crack may have the potential to clap and generate nonlinear outputs.

We expect to observe contact nonlinear phenomena, i.e. appearance of harmonics in the output recorded at the sensor point, for the models marked by the dashed rectangles in Fig. 4. That is, for the model with zero crack aperture, clapping will be guaranteed even when the model is excited using very small amplitudes, and thus harmonics of the output wave should be seen for all the three excitation amplitudes. While for the model with non-zero crack apertures, harmonics of the output wave are expected when the excitation amplitude is large enough to trigger the clapping between crack surfaces, i.e. the input excitation wave amplitude in terms of displacement is larger than or equal to the crack aperture size. Therefore, the model with crack aperture $a$ = 5.0e-5 mm should be able to generate harmonics when subjected to the excitation waves with amplitudes of both $A_{d_y}$ = 5.0e-5 mm and 5.0e-3 mm; for the model with crack aperture $a$ = 5.0e-3 mm, only the excitation wave with amplitude $A_{d_y}$ = 5.0e-3 mm can induce harmonics. Additionally, intact models with the same setup are simulated as references to the cracked models.

Fig. 5 presents an example of the time evolution of *y*-velocity within an excitation cycle *T* for both the intact and closed crack (i.e. $a$ = 0) models subjected to the excitation wave with the largest amplitude shown in Fig. 4 (i.e. $A_{d_y}$ = 5.0e-3 mm and $A_{v_y}$ = 3.1 m/s). Both models have reached steady state and the four time instances are respectively the beginning, first, second and third quarter of an excitation cycle. As the compressive wave continually excites at the bottom boundary, the wave gradually propagates into the model and moves upwards. A comparison between the two models clearly demonstrates how the crack perturbs the wave propagation. For the intact model (Fig. 5a), the wave fades out gradually as it propagates, with a small amount of reflection from the boundaries can be observed. While for the cracked model, when the wave reaches the crack, clapping between crack surfaces start to occur along the interface. The wave continually transmits through the crack, but with a distinct velocity reduction caused by the reflection of wave energy at the crack surfaces.

The simulation results of the models subjected to the three excitation waves are respectively presented in Fig. 6, Fig. 7 and Fig. 8, in terms of *y*-velocity at the sensor point and their corresponding FFTs. As can be seen from Fig. 6 that for the models excited using the amplitude of $A_{v_y}$ = 3.1e-4 m/s (i.e. $\varepsilon_y$ = 1.0e-8 and $A_{d_y}$ = 5.0e-7 mm), the output *y*-velocity at the sensor point for the intact model is still of perfect sinusoidal shape (Fig. 6a), as is indicated by the single peak generated at 100 kHz in its FFT (Fig. 6e). However, the output wave amplitude (~2.2e-4 m/s) is slightly smaller than the excitation amplitude (3.1e-4 m/s). This is mainly due to energy dissipation during the wave propagation. On the contrary, the output *y*-velocity at the sensor point for the closed crack model ($a$ = 0) shows a distinct distortion (Fig. 6b) and thus harmonics can be seen from its FFT (Fig. 6f). For the other two cracked models with apertures $a$ = 5.0e-5 mm and 5.0e-3 mm, as we expected earlier, no wave distortion or harmonics can be seen since the excitation amplitude ($A_{d_y}$ = 5.0e-7 mm) is too small to initiate clapping between crack surfaces.

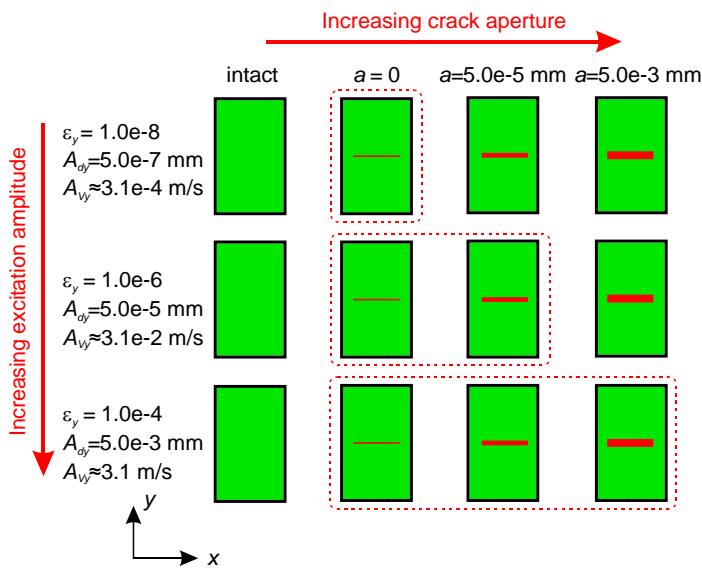

Fig. 4. FDEM models with increasing crack aperture and subjected to excitation waves with increasing amplitudes are created to investigate the relationship between crack aperture and nonlinear phenomena generation. Note the intact models are used for comparison purposes. The models located in the dashed rectangles are expected to generate nonlinear phenomena because of their potential of clapping between crack surfaces.

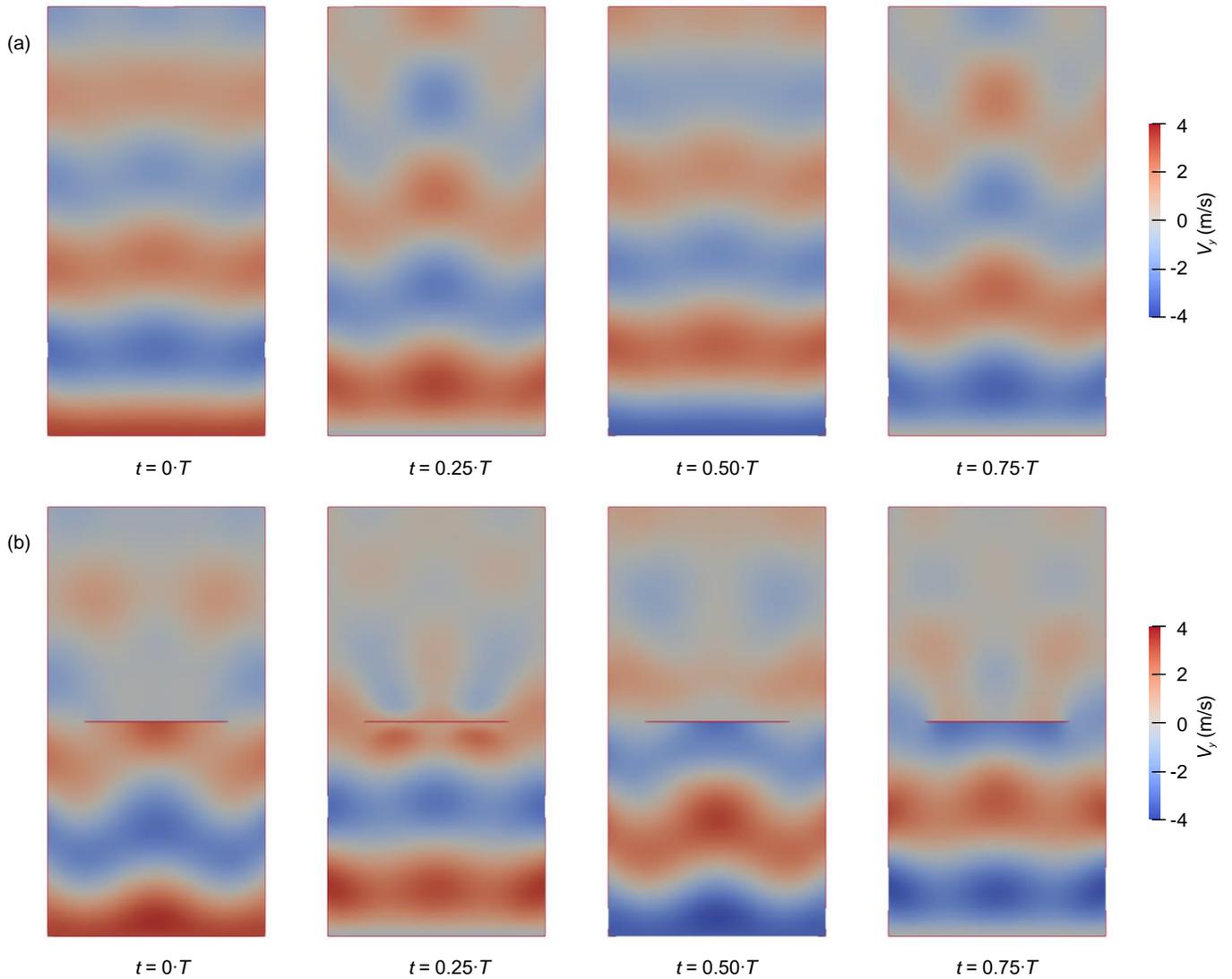

Fig. 5. Evolution of *y*-velocity within an excitation cycle *T* for (a) intact model and (b) model with a closed horizontal crack (i.e. *a* = 0) subjected to the excitation wave with the largest amplitude shown in Fig. 4 (i.e. $A_{d_y}$ = 5.0e-3 mm and $A_{v_y}$ = 3.1 m/s). The four time instances are respectively the beginning, first, second and third quarter of an excitation cycle.

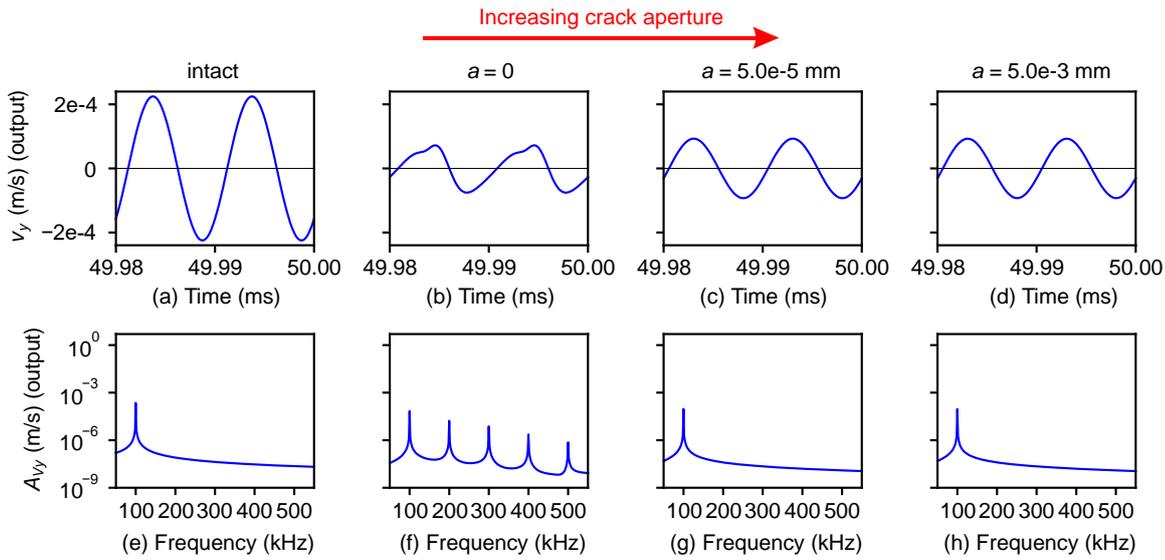

Fig. 6. Time evolutions of *y*-velocity at the sensor point (top row) and their FFTs (bottom row) for the intact model and cracked models (horizontal crack) with crack apertures *a* = 0 (closed crack), 5.0e-5 mm and 5.0e-3 mm, respectively. The models are subjected to compressive excitation wave with amplitudes $A_{v_y}$ = 3.1e-4 m/s (i.e. $\varepsilon_y$ = 1.0e-8 and $A_{d_y}$ = 5.0e-7 mm).

Similar results are obtained for the models subjected to the other two excitation amplitudes of $A_{v_y}$ = 3.1e-2 m/s (i.e. $\varepsilon_y$ = 1.0e-6 and $A_{d_y}$ = 5.0e-5 mm) and $A_{v_y}$ = 3.1 m/s (i.e. $\varepsilon_y$ = 1.0e-4 and $A_{d_y}$ = 5.0e-3 mm), as are shown in Fig. 7 and Fig. 8, respectively. As the excitation amplitudes increases to $A_{d_y}$ = 5.0e-5 mm, the model with crack aperture *a* = 5.0e-5 mm starts to clap and generate harmonics (Fig. 7c & g). While this excitation amplitude is still not sufficient to induce nonlinearity in the model with the largest crack aperture *a* = 5.0e-3 mm (Fig. 7h). As the excitation amplitude continually increases to $A_{d_y}$ = 5.0e-3 mm, i.e. reaches the value of the largest crack aperture, harmonics are visible in all three cracked models (Fig. 8f-h). When nonlinearity is observed in a specific model, the output wave generally displays distinct distortions, as can be seen from Fig. 6b, Fig. 7b-c and Fig. 8b-d, which is a result of wave perturbation by the clapping between crack surfaces. Additionally, all these simulations reveal that as long as the crack exists, the output wave amplitude will be significantly reduced compared with the intact model (first row of Fig. 6, Fig. 7 and Fig. 8, about half of the intact model), regardless of clapping and nonlinear elasticity generation. This is an evidence that the existence of crack indeed reduces the wave amplitude, as is mentioned earlier in Fig. 5b the velocity field of the cracked model.

To facilitate the comparison of these models and to summarize, FFTs of the *y*-velocity at the sensor point for the above simulations are collected in Fig. 9. We can observe that larger excitation amplitudes generally yields output waves with higher amplitudes and no nonlinear elasticity phenomenon are visible for the intact model. For the cracked models, nonlinear elasticity is generated only when the excitation amplitude is large enough to trigger the clapping between crack surfaces. All the simulations demonstrate that the FDEM is capable to simulate the expected behavior of crack induced nonlinear elasticity phenomenon.

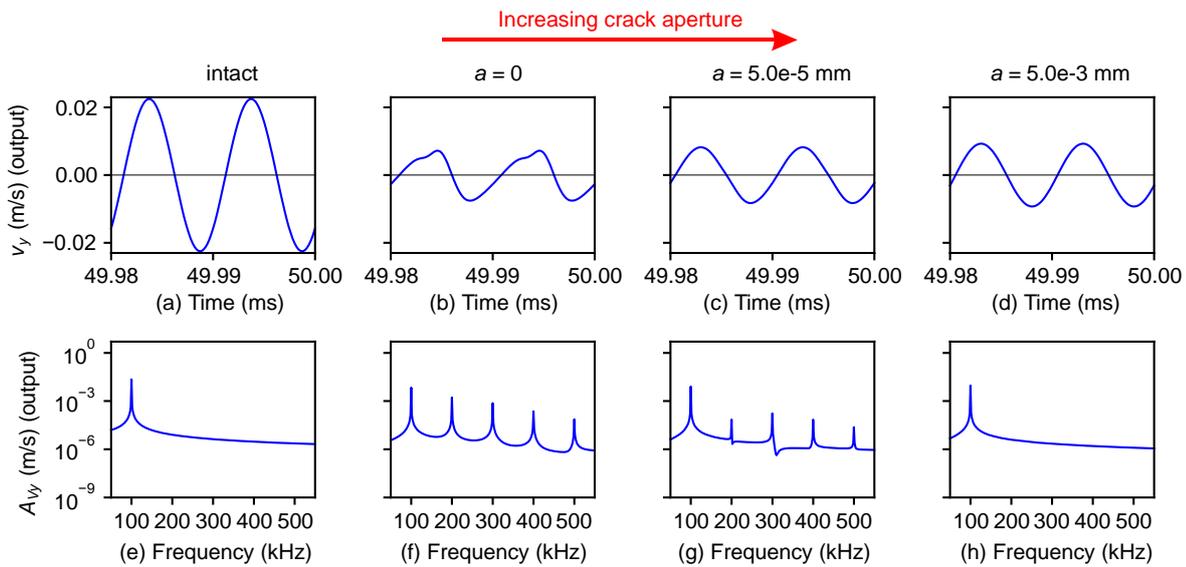

Fig. 7. Time evolutions of *y*-velocity at the sensor point (top row) and their FFTs (bottom row) for the intact model and cracked models (horizontal crack) with crack apertures $a = 0$ (closed crack), 5.0e-5 mm and 5.0e-3 mm, respectively. The models are subjected to compressive excitation wave with the amplitude $A_{v_y} = 3.1\text{e-}2$ m/s (i.e. $\varepsilon_y = 1.0\text{e-}6$ and $A_{d_y} = 5.0\text{e-}5$ mm).

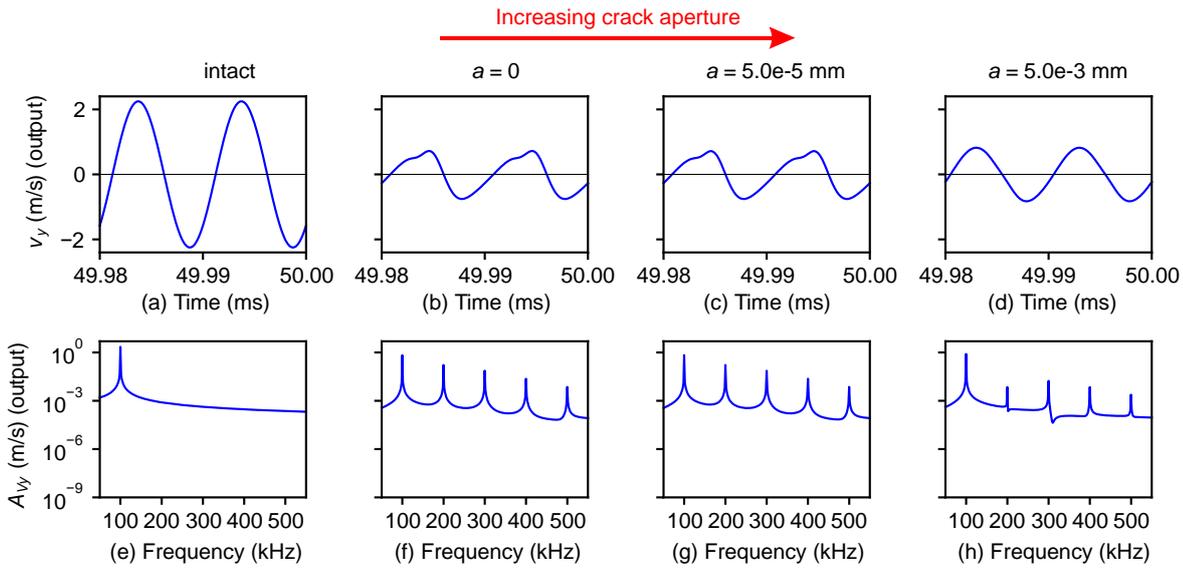

Fig. 8. Time evolutions of *y*-velocity at the sensor point (top row) and their FFTs (bottom row) for the intact model and cracked models (horizontal crack) with crack apertures *a* = 0 (closed crack), 5.0e-5 mm and 5.0e-3 mm, respectively. The models are subjected to compressive excitation wave with the amplitudes $A_{v_y}$ = 3.1 m/s (i.e. $\varepsilon_y$ = 1.0e-4 and $A_{d_y}$ = 5.0e-3 mm).

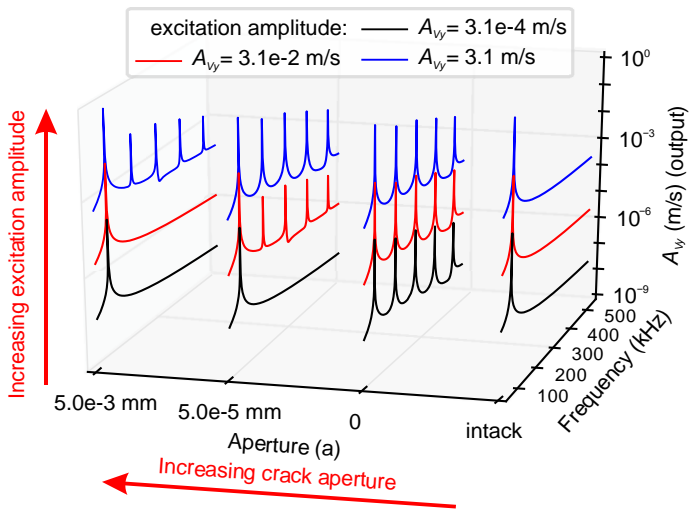

Fig. 9. Collection of the FFTs of *y*-velocity at the sensor point for all the simulation cases shown in Fig. 4.

However, for the model with crack aperture *a* = 5.0e-5 mm and subjected to excitation wave with amplitude $A_{d_y}$ = 5.0e-5 mm (Fig. 7c & g), and the model with crack aperture *a* = 5.0e-3 mm and subjected to excitation wave with amplitude $A_{d_y}$ = 5.0e-3 mm (Fig. 8d & h), since the excitation amplitudes is just about the same as the crack apertures, it seems only mild contacts between crack surfaces occur and the generated harmonics are not as strong and regular as their counterpart scenarios that having better contacts (e.g. Fig. 7b & f and Fig. 8c & g). In the next section, using the models with the crack aperture $a$ = 5.0e-5 mm, we have a closer look at the sensitivity of contact on nonlinear elasticity generation.

*3.3. Clapping and friction induced nonlinear phenomena*

To have a detailed check of the relationship between contact status along the crack interface and the overall nonlinear behavior, we choose another two excitation waves with amplitudes – $A_{d_y}$ = 2.5e-5 mm (i.e. $\varepsilon_y$ = 5.0e-7 and $A_{v_y}$ = 1.6e-2 m/s) and $A_{d_y}$ = 2.5e-4 mm (i.e. $\varepsilon_y$ = 5.0e-6 and $A_{v_y}$ = 1.6e-1 m/s) – one is slightly smaller and the other slightly larger than $A_{d_y}$ = 5.0e-5 mm that we have used earlier (Fig. 7c & g), to excite the model with horizontal crack and aperture $a$ = 5e-5 mm. Since the previous simulation shows that only mild contact occurs for this model subjected to excitation wave with amplitude $A_{d_y}$ = 5.0e-5 mm, the first excitation amplitude $A_{d_y}$ = 2.5e-5 mm may not be large enough to trigger contact between crack surfaces, whereas the second amplitude $A_{d_y}$ = 2.5e-4 mm may yield full contact along the crack interface. Therefore, these two new excitation waves, together with the one used earlier – $A_{d_y}$ = 5.0e-5 mm, form a scenario through which the sensitivity of contact on the nonlinearity phenomenon generation can be examined.

We tracked the changes of crack aperture (*a*) and relative tangential displacement (*s*) along the crack interface, and the overall normal ($F_N$) and tangential ($F_T$) contact force between crack surfaces, with respect to time. 15 pairs of monitoring points are placed evenly along the crack interface, and each pair of points are located symmetrically with respect to the crack axis. By calculating the relative position of each pair of monitoring points and resolving it into the directions perpendicular and parallel to the crack orientation, the change of crack aperture *a* and relative tangential displacement *s*, respectively, along the crack interface can be obtained. Here the positive *s* denotes that a pair of corresponding monitoring points has been moved clockwise compared with their original positions. The overall normal and tangential contact forces between crack surfaces are calculated by firstly resolving the normal and tangential contact forces between each contact element pair into the directions perpendicular and parallel to the crack orientation and then integrating them respectively along the crack interface. Since the overall normal and tangential contact forces are action and reaction forces, their absolute values are presented in the following analyses. The non-zero normal and tangential contact forces ($F_N > 0$ and/or $F_T > 0$) as well as the zero crack aperture ($a = 0$) are indicators of contact between crack surfaces.

The time evolution of crack aperture $a$, overall normal contact force $F_N$ and the FFT of $y$-velocity at the sensor point for the horizontal crack model ($a = 5.0\text{e-}5$ mm) subjected to the three excitation waves are presented in Fig. 10. In case of the small excitation amplitude $A_{d_y} = 2.5\text{e-}5$ mm, none of the 15 positions on the crack are in contact (i.e. $a > 0$ and $F_N = 0$). When the excitation amplitude $A_{d_y}$ increases to the value equals to the crack aperture, i.e. 5.0e-5 mm, some of the 15 positions on the crack are in contact. Since this is a partial contact along the crack, i.e. only a few positions along the crack are in contact, relatively small normal contact force is detected; however, harmonics are still visible. As the excitation amplitude continually grows to $A_{d_y} = 2.5\text{e-}4$ mm, the whole crack interface is in contact (i.e. $a = 0$ and $F_N > 0$). As a result, a relatively higher overall normal contact force $F_N$ together with the harmonics are generated. A comparison of the results shown in Fig. 10 also demonstrates that for the horizontal crack, the clapping at different positions along the crack interface generally occurs at the same pace. Note that here only the crack aperture $a$ and normal contact force $F_N$ are given since the relative tangential displacement $s$ and tangential contact force $F_T$ are negligible compared with the normal components (the tangential components are less than 2% of the normal ones).

The normal contact was primarily triggered in the horizontal crack model. Next, still use the three excitation waves and the same crack aperture $a = 5\text{e-}5$ mm, the inclined crack model presented in Fig. 3b is employed to investigate the sensitivity and contribution of both the normal and tangential contacts on nonlinear phenomenon generation. The results in terms of crack aperture $a$, relative tangential displacement $s$, overall normal ($F_N$) and tangential ($F_T$) contact force and FFT of $y$-velocity at the sensor point are presented in Fig. 11. Similar to the results of the horizontal crack model (Fig. 10), when the model is excited by the wave with small amplitude $A_{d_y} = 2.5\text{e-}5$ mm, no contact between the crack surfaces has occurred, and thus both normal and tangential contact forces are zero throughout and no harmonics are generated. For the intermediate excitation amplitude $A_{d_y} = 5.0\text{e-}5$ mm, again, contact only occurs at some positions along the crack interface. This is visible in the plots of crack aperture and contact forces. As only partial contact happens, very small normal and tangential contact forces are induced. Nevertheless, nonlinear features of the output wave are observed, as is evidenced by the presence of harmonic frequencies. In case of the largest amplitude $A_{d_y} = 2.5\text{e-}4$ mm, clap occurs along the entire crack interface, and higher normal and tangential contact forces are generated, together with a larger oscillation amplitude of both the crack aperture and shear displacement. In addition to the crack aperture change, distortion (asymmetry) is also clearly visible in the relative tangential displacement, which is due to the presence of tangential contact force that reduces the relative movement of crack faces in the tangential direction. While for the contact force, the overall tangential contact force is smaller than the normal contact force, as the former can never exceed the latter, i.e. $F_T \leq \mu F_N$, where $\mu = 1$ in current simulations.

On the contrary to the horizontal crack model (Fig. 10), contacts at the 15 positions along the crack interface in the inclined crack model do not occur at the same pace (Fig. 11a & b), which can also be evidenced by the apparent fluctuation and longer occurrence time range in the contact force (Fig. 11c). This is due to the different arriving time of the wave to the crack surface. Additionally, since the effect of excitation wave on local particle vibration along the inclined crack interface has to be resolved into directions perpendicular and parallel to the crack orientation to obtain the normal and tangential response, the magnitude of the resulting crack aperture and normal contact force change (Fig. 11a & c) for the inclined crack model is smaller than that of the horizontal one (Fig. 10a & b). Moreover, the harmonics generated for both the horizontal and inclined crack models subjected to the intermediate excitation wave ($A_{d_y}$ = 5.0e-5 mm) due to partial contact demonstrate that the nonlinear elasticity is very sensitive to the crack surface contact and FDEM has the capability to capture such sensitivity.

As a way to further demonstrate which part of the crack has experienced the most nonlinearity, we have calculated the FFTs of the crack aperture change at the 15 positions along the crack interface for both the horizontal and inclined crack models subjected to the excitation wave with the largest amplitude $A_{d_y}$ = 2.5e-4 mm. The FFTs of the relative tangential displacement change at the 15 positions for the inclined crack model are also calculated. The results for the horizontal and inclined crack models are presented in Fig. 12 and Fig. 13, respectively. Fig. 12 illustrates that for the horizontal crack model, the induced nonlinear elasticity distributes uniformly along the crack interface. While for the inclined crack model, pronounced nonlinear behaviors induced by normal displacement (crack aperture change) mainly occur on the slightly left hand side of the crack (Fig. 13a, negative positions). This is due to the fact that the compressive wave will first reach the left hand side of the crack; while since it takes longer time to reach the right hand side, the wave amplitude will be reduced because of energy dissipation. This results in a larger normal vibration on the left hand side of the crack and hence, more dominant of the nonlinear elasticity generation. The harmonics of shear displacement show the opposite trend, i.e. more pronounced nonlinear behaviors are located on the right hand side of the crack (Fig. 13b, positive positions). This is probably because that as the compressive wave propagates to the crack, part of it turns into shear wave and moves upwards along the crack interface. Finally, all of the converted shear waves accumulate at the right hand side corner of the crack and result in a higher tangential vibration. Further simulations will be conducted to explore the influence of crack orientation on nonlinear elasticity generation in the near future.

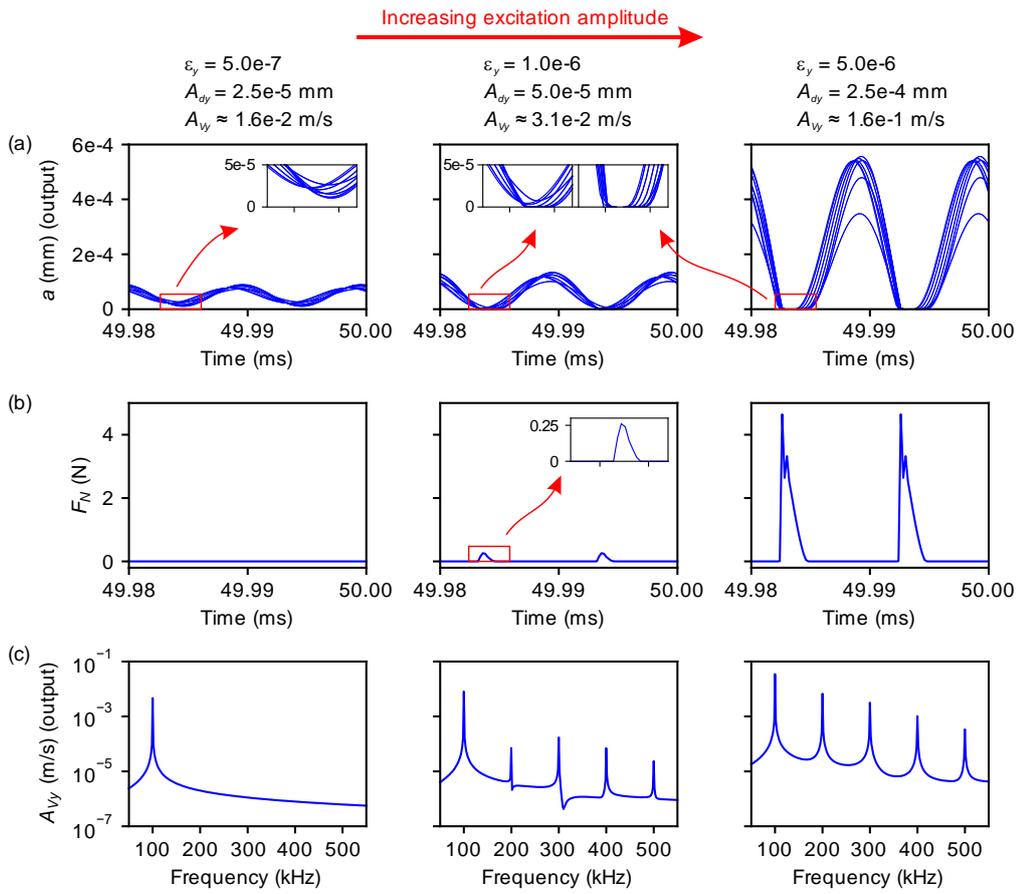

Fig. 10. Response of the model with horizontal crack (*a* = 5.0e-5 mm) and subjected to excitation waves with amplitude of $A_{d_y}$ = 2.5e-5 mm, 5.0e-5 mm and 2.5e-4 mm, respectively: (a) time evolution of aperture *a* at the 15 positions on the crack interface; (b) time evolution of overall normal contact force $F_N$ between crack surfaces; (c) FFT of *y*-velocity at the sensor point. The insets give a zoom-in of the results at specific time range.

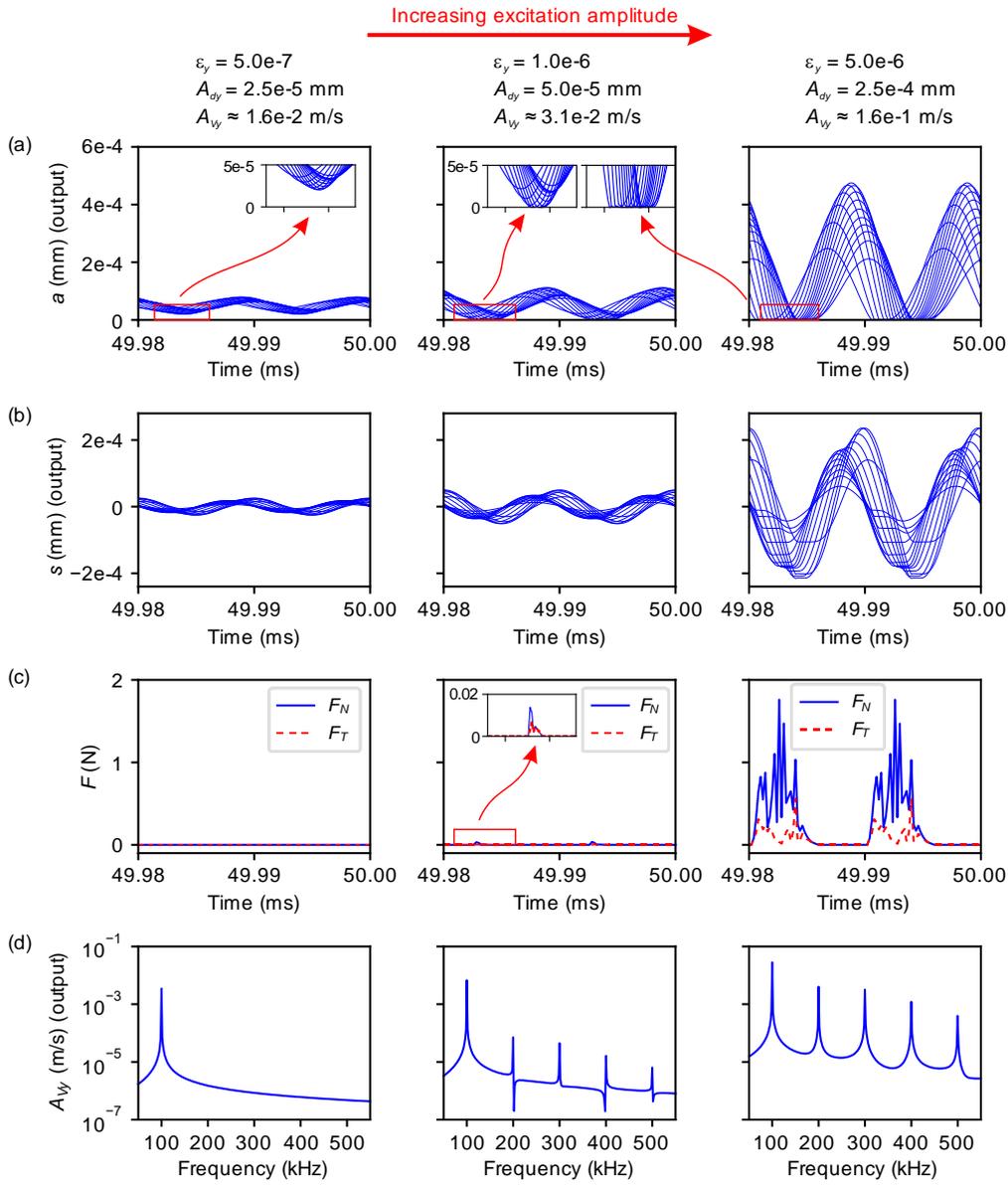

Fig. 11. Response of the model with 30º inclined crack ($a$ = 5.0e-5 mm) and subjected to excitation waves with amplitude of $A_{d_y}$ = 2.5e-5 mm, 5.0e-5 mm and 2.5e-4 mm, respectively: (a) time evolution of aperture $a$ at the 15 positions on the crack interface; (b) time evolution of relative tangential displacement $s$ at the 15 positions on the crack interface; (c) time evolution of overall normal ($F_N$) and tangential ($F_T$) contact forces between crack surfaces; (d) FFT of $y$-velocity at the sensor point. The insets give a zoom-in of the results at specific time range.

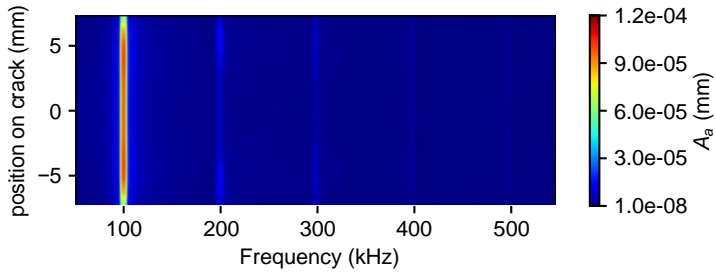

Fig. 12. FFTs of crack aperture change at the 15 positions on the crack interface for the model with horizontal crack ($a$ = 5.0e-5 mm) and subjected to the excitation wave with amplitude $A_{d_y}$ = 2.5e-4 mm.

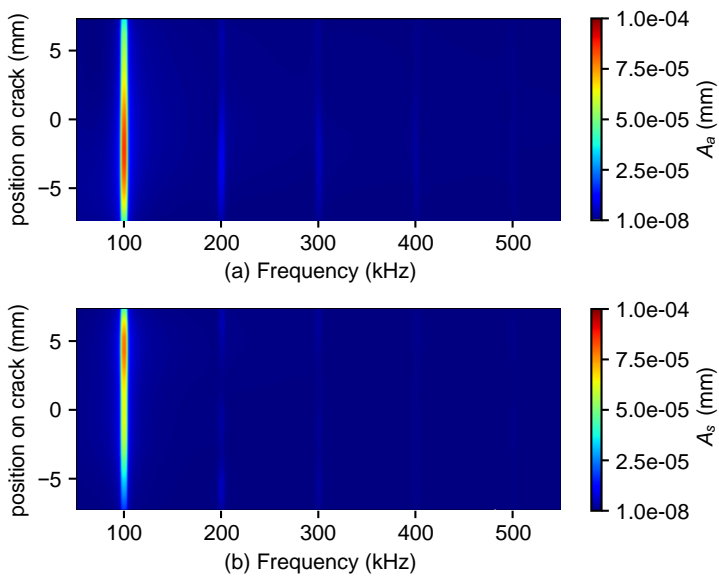

Fig. 13. FFTs of (a) crack aperture and (b) relative shear displacement change at the 15 points on the crack interface for the model with 30º inclined crack ($a$ = 5.0e-5 mm) and subjected to the excitation wave with amplitude $A_{d_y}$ = 2.5e-4 mm.

## 4. Conclusions

In this paper we have introduced FDEM by describing the application to an intact solid and two simple crack models – one normal to the forcing wave and one with crack at a 30º angle to the wave. In the FDEM model, the solid is discretized into finite elements to capture the wave propagation in the bulk material, and the finite elements along the two sides of the crack also behave as discrete elements to track the normal and tangential interactions between crack surfaces. By employing FDEM, the cracked solid can be explicitly realized and particularly, the contacts along the sides of the crack can be uniformly processed using well-developed DEM-based algorithms.

We first create a series of simulation scenarios in which different combinations of crack aperture and excitation amplitude for the model with horizontal crack to examine the capability of FDEM for nonlinear elasticity simulation. The simulation results reveal that larger excitation amplitudes generally yield output waves with higher amplitudes. For the intact model, the output wave is harmonically oscillating at the fundamental frequency and no nonlinear elasticity phenomenon has been observed. Whereas for the cracked models, nonlinear elasticity is generated when the excitation amplitude is sufficient to trigger the contact between crack surfaces, which is evidenced by the appearance of harmonics of the output wave. The simulations demonstrate that FDEM is capable to simulate the expected behavior of crack induced nonlinear elasticity.

Then for the models with horizontal and 30º inclined cracks of aperture $a$ = 5e-5 mm, we have a detailed check of the relationship between contact status along the crack interface and the overall nonlinear behavior. It is shown that for the horizontal crack model, the nonlinear elasticity is mainly triggered by the normal contact, and it is distributed uniformly along the crack interface. Whereas for the inclined crack model, both the normal and tangential contacts contribute to the resulting nonlinear elasticity generation: pronounced nonlinear behaviors induced by normal displacement mainly occur on the slightly left hand side of the crack; the harmonics of shear displacement show the opposite trend, i.e. more pronounced nonlinear behaviors are located on the right hand side of the crack. Additionally, the harmonics generated for both the horizontal and inclined crack models due to partial contact demonstrate that the nonlinear elasticity is very sensitive to the crack surface contact. The simulations not only reveal the influence of normal and tangential contact on the nonlinear elasticity generation, but also demonstrate the capabilities of FDEM for revealing the causality of nonlinear elasticity in cracked solid and its potential to assist in Non-Destructive Testing (NDT).

The simulation results are for two simple and idealized systems. Although in the current simulation only flat crack surfaces are used, because of the approach in FDEM for contact processing, FDEM has the capability of considering rough cracks by either explicitly creating crack surfaces with certain waviness or employing an empirical crack constitutive model that implicitly considers crack roughness for contact interaction calculation, as is already implemented and tested earlier in FDEM [59]. Importantly, FDEM can also simulate the cracks with various shapes (open, closed, flat, bending curve), the interaction between different types of defects as well as crack propagation under loads. Future investigations considering complex crack geometries will be conducted on the influence of these factors on nonlinear phenomenon generation in defected solid using FDEM.

## Acknowledgments

The Department of Energy Office of Basic Energy Research, Geoscience, supported this work. Technical support and computational resources from the Los Alamos National Laboratory Institutional Computing Program are highly appreciated.